\documentclass[aps,prl,amsfonts,amsmath,twocolumn,groupaddress]{revtex4-1}
\pdfoutput=1 
\usepackage{graphicx}
\usepackage{amssymb}
\usepackage{amsmath}
\usepackage{amsfonts}
\usepackage[usenames]{color}
\usepackage{gensymb}
\usepackage{amsbsy}
\usepackage{epsfig}
\usepackage{color}
\newcommand{\dif}{\mathrm{d}}

\newcommand{\green}{}
\begin{document}


\title{Three Body Fluctuation-Induced Interaction at Fluid Interfaces:\\
A Strong Deviation from the Pairwise Summation}


\author{Ehsan Noruzifar}
\author{Jef Wagner}
\author{Roya Zandi}
\affiliation{Department of Physics and Astronomy, 
University of California, Riverside, California 92521, USA}


\date{\today}

\begin{abstract}
We present a new method based on the scattering technique to
investigate fluctuation-induced forces at a fluid interface. 
The scattering approach, well suited to the study of many body systems 
of arbitrary geometries, is augmented to include boundary fluctuations. 
Using this method, we study the deviation of the total fluctuation-induced
interaction from the sum of pairwise energies for three colloidal particles.  
We consider both frozen and fluctuating colloids and obtain a very good agreement between analytical and numerical 
results. We find a marked difference in the three body fluctuation-induced 
free energy between the frozen and fluctuating colloids, both in sign and relative size.
\end{abstract}


\maketitle

With the construction of micro- and nano-eletro-mechanical devices
(MEMS and NEMS) \cite{serry,buks,chan} and the desire to understand biological processes in greater details, 
it has become necessary to have a
complete understanding of fluctuation-induced forces. These forces
were first predicted by Casimir in 1948 who showed that confinement of
zero point energy fluctuations of the electromagnetic field between
two conducting plates in vacuum gives rise to long ranged attractive
interactions \cite{casimir,casimir2}.  Thirty years later, Fisher and de Gennes noted that
imposing constraints on the thermal, rather than quantum,
fluctuations in fluids in the vicinity of their critical points also
results in similar long-ranged forces \cite{fisher}.  Since then, the thermal
versions of Casimir forces have been investigated in various
theoretical and experimental studies \cite{zandi,garcia,bruinsma,diet,hertlein1,kardar_rods, mine_epje, mine_pre, oettel06, oettel07,
  eft_deserno1, eft_deserno2,lin}.

In particular, due to the industrial applications and importance in
biological systems, the thermal fluctuation-induced forces between
colloidal particles have been studied 
within the last few years. For example, the experiments of Hertlein {\em et. al.}
reveal the presence of repulsive and attractive fluctuation-induced
forces on colloidal particles immersed in a near-critical binary
mixture in the vicinity of a wall \cite{hertlein1}.  The thermal
fluctuation-induced force also plays an important role in the
effective interaction between colloidal particles trapped between two
fluid phases \cite{oettelrev1,oettelrev2}.

There have been previous works using various techniques to study
fluctuation-induced forces at interfaces with interesting
results \cite{kardar_rods, mine_epje, mine_pre, oettel06, oettel07,lin,
  eft_deserno1, eft_deserno2}.  There are, however, a number of
complications associated with each of these methods. The most common
is the difficulty of calculating the interaction between colloids at
very small separations and matching it with the Derjaguin
approximation \cite{derjaguin}. Moreover, extending the previous methods to various geometries
and many particle systems can be tedious.

In this work we study the interaction between colloids trapped at a
fluid interface using the scattering formalism
\cite{scattering_letter, scattering_scalar, scattering_vector}, which
is widely used to obtain quantum fluctuation-induced forces. The scattering formalism presents a straightforward
approach for treating various geometries and many body 
systems and is numerically faster compared to previous methods. 

Furthermore, the scattering method simplifies the fluctuation-induced
calculations by writing 
the {energy} into the translation
matrices ($\mathbb{U}$) and scattering matrices or $T$-matrices
($\mathbb{T}$). The {translation and $T$-}matrices {describe how} a fluctuation
propagates through the field between the objects and {how}
fluctuations interacts with the objects, respectively.

The important difference between the QED Casimir effect and the fluctuation-induced forces
between colloids at an interface is that the
colloids and their boundaries can fluctuate.  This work extends the
scattering method to include these effects, by separating out the
interface and colloid Hamiltonians. This new extension can 
simplify the problem by separating the properties of the colloid,
as well as the type of fluctuations, from the
calculation of the $T$-matrix.
Note, in contrast to previous works \cite{kardar_rods, mine_epje, mine_pre, oettel06, oettel07,lin}, 
the calculation of the T- and the translation matrices are completely decoupled.

In order to demonstrate the effectiveness of this new technique, we
calculate the fluctuation-induced forces between three spherical Janus
colloids at an interface. We study three different fluctuation scenarios: (i) colloids frozen at the interface; 
(ii) bobbing colloids that fluctuate only
vertically; and (iii) bobbing and tilting colloids that
both fluctuate vertically and tilt side to side. Surprisingly we find
different three body behaviors for fixed and fluctuating colloids.  The three body
effect for {the} fixed colloids is repulsive and comparable to the two body interaction, while 
for fluctuating colloids is attractive and comparable 
only at short separations.

In the following, we briefly explain the derivation of the scattering
formalism for colloids at fluid interfaces. Moreover, we show some numerical and
analytical results obtained with this technique.  Note that due to the
rapid communications format, we only outline the important steps. The
details of the derivations and calculations are left for a more
complete exposition \cite{inprep}.

We consider an infinite interface between two fluid phases
characterized by the surface tension $\sigma$.  At equilibrium, the
interface is flat and placed at $z=0$.  We suppose that the thermal height fluctuations 
of the interface are small without overhanges
and bubbles. Therefore, we express the interface height profile in the
Monge representation, $\it{i.e.}$ $z=u({\bf x})$. 
The capillary wave
Hamiltonian \cite{oettel06, oettel07} then becomes
\begin{equation}
  \label{eq:interface-hamiltonian}
  \mathcal{H}_{\text{int}}[u] \approx
  \frac{\sigma}{2}\int_{\mathbb{R}^2} \dif^2 x\,\left[  
    \big( \nabla u\big)^2 + \frac{u^2}{\lambda_{\rm c}^2}\right]\,,
\end{equation}
where $\lambda_{\rm c} = \sqrt{{\sigma}/{(\Delta \rho g})}$ is the
capillary length with $\Delta\rho = \rho_1 - \rho_2$, the difference
in the mass density of the two fluid phases.

We now introduce $N$ colloidal particles at the interface between two
fluid phases.  The colloid Hamiltonian will
be defined as the energy costs associated with the insertion of
a colloid,
\begin{multline}
  \label{eq:colloid_hamiltonian}
  \mathcal{H}_{\text{col}}^{i} [f_i,h_i] = 
  -\frac{\sigma}{2} \int_{\Omega_i} \dif^2 x
  \left[\big(\nabla f_i\big)^2+\frac{f^2}{\lambda_{\rm c}^2}\right] \\
  +\sigma {\Delta\Omega_i}[f_i] 
  + (\sigma_{i,\text{I}} - \sigma_{i,\text{II}}) 
  \Delta A_{i,\text{I}}[f_i,h_i]\,,
\end{multline}
where $h_i$ is the height of the center of mass of the colloid $i$
with respect to the equilibrium interface, and
$\sigma_{i,\text{I(II)}}$ is the surface tension between the colloid
$i$ and the fluid phase I(II).  The contact line field $f_i({\bf x})$
is the height at which the fluid interface intersects the colloid,
and is extended into the interior of the colloid. The cross sections
of the colloid $i$ with the equilibrium and fluctuating interfaces are
denoted by $\Omega_i^{\rm ref}$ and $\Omega_i$, respectively, and
$\Delta\Omega = \Omega_{\rm ref}-\Omega$ is the change in the
projected area.  The surface area of the colloid in contact with the
fluid phase I(II) is represented by $A_{\rm I(II)}$ where $\Delta
A_{i,\text{I}}$ is the change in the area of the colloid $i$ in the
fluid phase I. Note that for Janus particles the total area of the colloid does
not change $\Delta A_{\rm I}=-\Delta A_{\rm II}$. In addition, the
gravitational energy due to the colloid height $h_i$ is negligible and
can be ignored in Eq.~\eqref{eq:colloid_hamiltonian} because
$h_i/\lambda_{\rm c} \lll 1$.

The first term in Eq.~\eqref{eq:colloid_hamiltonian} is associated
with excluding the projected areas $\Omega_i$ from the interface since
the integration domain in Eq.~\eqref{eq:interface-hamiltonian} 
covers the whole ${\mathbb R}^2$ space. The second term represents the
energy costs related to the change of the projected area for a colloid
in the fluctuating vs reference interface, and the third term is the
energy costs due to the change in the area of the colloids in the
phase I and II.

The total Hamiltonian of the system is the sum of the interface and
$N$ colloid Hamiltonians, $ \mathcal{H}_{\text{tot}} =
\mathcal{H}_{\rm int}[u] + \sum_{i=1}^N
\mathcal{H}_{\text{col}}^i[f_i,h_i] \, $, and the partition function
$\mathcal Z$ is the sum over all interface and colloid configurations
\begin{equation}
 \label{eq:partition-function}
 \mathcal{Z} = \int_{\mathcal C} \mathcal{D} u \prod_{i=1}^N 
 \mathcal{D} f_i \exp \left[-\frac{{\mathcal H}_{\text{tot}}}{k_B T} \right]\,,
\end{equation}
with $\mathcal C$ representing the constraints imposed by the
colloidal particles on the interface height fluctuations. At the
surface of the colloid $i$ the interface height field $u$ is
constrained by the contact line field $f_i$. A Dirac delta functional
is used to implement the constraints, $\delta[u - f_i] = \int
\prod_{i=1}^N\mathcal{D} \psi_i \exp [ \imath \int_{\Omega_i} \dif^2
  x\, \psi (u -f_i) ]$, where we have introduced the auxiliary field
$\psi_i({\bf x})$. Equation~\eqref{eq:partition-function} can be
separated as
\begin{equation}
  \label{eq:partition_int_col}
  \mathcal{Z}= \int \prod_{i=1}^N{\mathcal D}\psi_i \,{\mathcal Z}_{\rm int}\,
  {\mathcal Z}_{\rm col}\,,
\end{equation}
where ${\mathcal Z}_\text{int}$ and ${\mathcal Z}_\text{col}$
correspond to the partition functions related to the interface and
colloid fluctuations respectively.  

The integration over the scalar field $u$ in the interface partition
function can be performed and results in
\begin{equation}
  \label{eq:partition-int-gij}
  \mathcal{Z}_{\text{int}} = C_0\, 
  \exp\big(-\frac{k_B T}{2\sigma}\sum_{i,j=1}^N G_{ij}\big)\,,
  \end{equation}
where $C_0$ is a constant and 
\begin{equation}
 \label{eq:Gij} 
 G_{ij} = \int_{\Omega_i} \dif^2{x} \int_{\Omega_j}\dif^2{x}' \psi_i({\bf x})\, G({\bf x}, {\bf x}')\,\psi_j({\bf x}')\,.
\end{equation}
In Eq.~\eqref{eq:Gij}, $G({\bf x},{\bf x}')$ is the Green's function of the capillary
wave equation for the free interface and can be expanded in terms of the solutions to the
capillary wave equation, {\it i.e.} $(-\nabla^2 +\lambda_{\rm c}^2)u({\bf
  r})=0$. Using separation of variables in a coordinate system
centered on a colloid, the capillary wave equation has two solutions:
the solution regular at origin, which corresponds to the incident
field in the scattering method $\phi_\alpha^{\rm inc}({\bf r})$; and a
solution that dies off at infinity, which corresponds to the scattered
field $\phi^{\rm sct}_m({\bf r})$. The expansion of the Green's
function $G({\bf x},{\bf x}')$ and the auxiliary field $\psi_i$ in
terms of these solutions are $G({\bf x},{\bf x}') = \sum_\alpha
c_\alpha \,\phi_\alpha^{\rm inc}({\bf x}) \phi_\alpha^{\rm sct}({\bf
  x}')$ and $\psi_i({\bf x}) = \sum_\alpha \Psi_\alpha
\phi_\alpha^{\rm inc}({\bf x})$, respectively, with $c_\alpha$ the
expansion coefficient and $\Psi_\alpha$ the multipole coefficient.
Plugging these equations into Eq.~\eqref{eq:Gij}, we find
$G_{ij}=\sum_{\alpha\beta} c_\alpha \Psi_{i\alpha}{\mathbb
  U}_{\alpha\beta}^{ij} \Psi_{j\beta}$ for $i \ne j$ and
$G_{ii}=-\sum_{\alpha\beta} c_\beta \Psi_{i\alpha} [{\mathbb
    T^i}_{\alpha\beta}]^{-1} \Psi_{i\beta}$ for $i =j$. The
translation matrix ${\mathbb U}^{ij}$ describes the couplings between
the partial waves on distinct colloids. The scattering amplitude
matrix ${\mathbb T^i}$ relates the incident and scattered fields, and
depends only on the shape of the colloid. Inserting the expressions
for $G_{ij}$ in Eq.~\eqref{eq:partition_int_col}, we find
\begin{multline}
 \label{eq:Z_int_multipoles}
 \mathcal{Z}_{\rm int} = C_0\,\exp\bigg[ \frac{k_B T}{2\sigma}
  \sum_{i=1}^N {\bf\Psi}_i^{\dagger} {\bf C} \, 
      [{\mathbb T}^{i}]^{-1} {\bf\Psi}_i \\
      -\frac{k_B T}{2\sigma}\sum_{i\ne j=1}^N 
      {\bf \Psi}_i^\dagger\, {\mathbf C}\,{\mathbb U}^{ij}\, {\bf \Psi}_j
      \bigg]\,.
\end{multline}

The matrix ({$\bf C$}) in Eq.~\eqref{eq:Z_int_multipoles} is a diagonal matrix with the elements that are the
coefficient of the Green's function expansion, and ${\bf \Psi}_i$ is a
vector with elements that are the multipole moments associated with
the particle $i$.
 
We evaluate the colloid partition function following the same steps
as the ones we took to calculate the interface partition function. The colloid Hamiltonian in
Eq.~\eqref{eq:colloid_hamiltonian} is expanded to be quadratic in the
contact line field $f_i$, and then the field is expanded in
terms of the incident solutions, 
\begin{equation}
  \label{eq:contact_line_expansion}
  f_i({\bf x}) = \sum_\alpha P_\alpha
  \phi_\alpha^{\rm inc}({\bf x}). 
\end{equation}
After expansion the colloid partition function is
\begin{multline}
  \label{eq:Z_col_multipoles}
  \mathcal{Z}_{\rm col}=\int \prod_{i=1}^N {\mathcal D}{P}_i 
  \exp\bigg\{-\frac {1} {k_BT}\sum_{i=1}^N {\bf P}_i^{\rm T}
  \,{\bf H}_{{\rm col}}^i\,{\bf P}_i \\
  -\frac{\imath}{2}\,({\bf P}_i^{\dagger}{\bf\Psi}_i
  +{\bf P}_i{\bf\Psi}_i^{\dagger})\bigg\}\,,
\end{multline}
where the exact form of ${\bf H}_{{\rm col}}^i$ depends on the type 
of colloid  fluctuations and will be presented below.  We now
insert Eqs.~\eqref{eq:Z_col_multipoles} and
\eqref{eq:Z_int_multipoles} into Eq.~\eqref{eq:partition_int_col} to
obtain the total partition function ${\mathcal Z}$. The resulting
partition function is a Gaussian integral over the multipoles $\bf \Psi$ and $\bf P$. The
fluctuation-induced free energy at temperature $T$ can be then
obtained through ${\mathcal E}/k_B
T=-\ln\left({\mathcal{Z}}/{{\mathcal Z}_\infty}\right)$, with
${\mathcal Z}_\infty$ corresponding to the partition function with all
the colloids placed at infinite distance from each other.  The
fluctuation-induced energy as a function of the $T$-matrix and the
translation matrix $U$ then reads as
\begin{equation}
 \label{eq:energy-final}
 \frac{\mathcal E}{k_BT}=\frac{1}{2}\ln{\rm det}({\mathbf 1}
 -\widetilde{\mathbf T}{\mathbf U})\,,
\end{equation}
where ${\mathbf U}_{ij}={\mathbb U}^{ij}(1-\delta_{ij})$ and
$\widetilde{\mathbf T}_{ij}=\widetilde{\mathbb T}^{i}\delta_{ij}$ with
$\widetilde{\mathbb T}^i$ the modified T-matrix of the colloid $i$
given by
\begin{equation}
  \label{eq:tmatrix-modified}
  \widetilde{\mathbb T}^i = {\mathbb T}^i - {\mathbb T}^i 
  \big[{\mathbf C}\,{\mathbb T}^i\,{\mathbf C}^{-1}+
    {2\over\sigma}{\mathbf C}\,
    {\mathbf H}_{{\rm col}}^i\big]^{-1}{\mathbb T}^i\,.
\end{equation}
Note while ${\mathbb T}$ depends only on the shape of
the colloid, $\widetilde{\mathbb T}$ also depends on ${\mathbf H}_{{\rm col}}$
which in turn depends on the specific fluctuations that the colloid
undergoes. Equations \eqref{eq:energy-final} and
~\eqref{eq:tmatrix-modified} can be used to obtain the
fluctuation-induced interaction between N colloidal particles at a
fluid interface.

We now employ Eq.~\eqref{eq:energy-final} to calculate the
fluctuation-induced free energy between three identical spherical
colloids with radius $R$ trapped at an interface. For simplicity we 
assume that the colloids are Janus particles, {\it i.e.} the contact
line is pinned to the colloid surface such that the area of the
colloidal particles in two fluid phases does not change, {\it i.e.} $\Delta
A_{i,{\rm I}}=0$. In addition, we set the equilibrium contact angle to
$\pi/2$.  Using the solutions to the Helmholtz
equation in polar coordinates, the T-matrix for colloidal particles with circular cross section at the
flat interface reads
 \begin{equation}
 \label{eq:t-matrix_sphere}
  {\mathbb T}_{m}=-\frac{I_m(R/\lambda_{\rm c})}{K_m(R/\lambda_{\rm c})}\,,
 \end{equation}
where $I_m$ and $K_m$ are the modified Bessel functions of the first
kind and second kind, respectively. In order to obtain the fluctuation-induced energy given by Eq.~\eqref{eq:energy-final}, 
it is necessary to calculate ${\mathbf H}_{{\rm col}}$ or more specifically the 
modified $\widetilde{\mathbf T}$ given in Eq.~\ref{eq:tmatrix-modified}. To find ${\mathbf H}_{{\rm col}}$ we assume that
the colloids
are: (i) frozen, allowing no fluctuation; (ii) bobbing, allowing
the colloids to fluctuate up and down; and (iii) bobbing and tilting,
allowing the colloids to fluctuate both up and down as well as tilt
side to side.

For a frozen colloid at the interface, the colloidal particles are
exposed to an infinite potential barrier to fluctuate, therefore the
elements of the matrix ${\bf H}_{{\rm col},i}$ tend to infinity and
from Eq.~\eqref{eq:tmatrix-modified} $\widetilde{\mathbb T}={\mathbb
  T}$. For bobbing colloidal particles, the projected area does not
change and therefore $\Delta\Omega = 0$ in
Eq.~\eqref{eq:colloid_hamiltonian}.  Substituting 
Eq.~\eqref{eq:contact_line_expansion} into
Eq.~\eqref{eq:colloid_hamiltonian}, we can calculate the matrix
${\mathbf H}_\text{col}$.  In the limit $R/\lambda_{\rm c} \ll 1$, we
find ${\mathbf H}_{\rm col}\approx -\pi\sigma R^2/\lambda_{\rm c}$ for
$m=0$ and ${\mathbf H}_{\rm col}\approx\infty$ for $|m|>0$. From Eq.~\eqref{eq:tmatrix-modified}
we immediately find that for $m=0$, $\widetilde{\mathbb T}_0 \approx
0$ and for $m\ne 0$, $\widetilde{\mathbb T}_m \approx {\mathbb T}_m$.

For colloids that both bob and tilt, the projected area changes with
fluctuations, {\it i.e.} $\Delta \Omega \ne 0$.  After calculating the
change in the projected area for small tilt angles, using
Eq.~\eqref{eq:contact_line_expansion}  in the
limit $R/\lambda_{\rm c} \ll 1$ we find ${\mathbf H}_{\rm col}\approx
-\pi\sigma R^2/\lambda_{\rm c}^2, 0, \infty $ for $m=0$, $|m|=1$ and
$|m|> 1$ respectively.  Therefore, we have $\widetilde{\mathbf T}_0,
\widetilde{\mathbf T}_{\pm 1} \approx 0$, for $|m|=0,1$ and
$\widetilde{\mathbb T}_m \approx {\mathbb T}_m$ otherwise. 

To
calculate the fluctuation-induced energy given in
Eq.~\eqref{eq:energy-final}, we also need to calculate the translation
matrix in polar coordinate, which can be obtained by using the Graff's
addition theorem for Bessel functions \cite{abram64}.  We find
${\mathbb U}^{ij}_{mm'} = (-1)^{m'}
e^{\imath(m-m')\phi_{ij}}K_{m-m'}(d/\lambda_{\rm c})$ with $\phi_{ij}$
the angle between the coordinate systems fixed to the colloids $i$ and
$j$.

Employing the translation and $T$-matrices described above, 
we write
\begin{figure}
  \begin{center}
    \includegraphics[scale=.8]{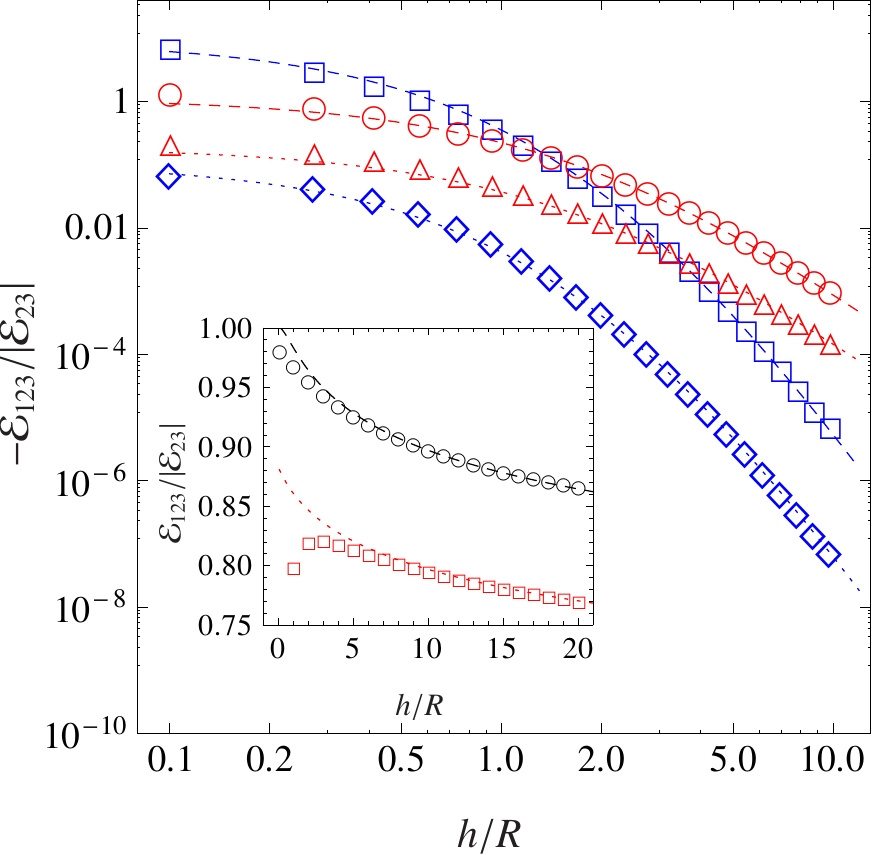}%
    \caption{\label{fig:3partfixed} (Color online) 
    Three body effect for colloids on a line with $d_{12}=d_{13}=d_{23}/2$
    and on an equilateral triangle with $d_{12}=d_{13}=d_{23}$:
    The ratio of the three body energy ${\mathcal E}_{123}$ 
    to the absolute value of the two body energy
      ${\mathcal E}_{23}$ vs. the surface-to-surface separation of
      the colloids 1 and 2, $h/R$.  
      The dashed and dotted lines represent the asymptotic results of Eqs.~\eqref{eq:3-bobing-asym} and \eqref{eq:3-tilt-asym} for the colloids 
      in the linear and triangular configurations, respectively. 
      Blue squares (linear configuration) and diamonds (triangular configuration) denote the numerical results for the bobbing and tilting colloids. 
      Red circles (linear configuration) and triangles (triangular configuration) show the numerical results for the bobbing only colloids. 
      Inset: Three body effect for {\em fixed} colloids 
      in the linear and triangular configurations. The dashed and dotted lines show the 
      asymptotic result in Eq.~\eqref{eq:three-fixed}. The black circles (linear configuration) and 
      red squares (triangular configuration) represent the numerical results for the fixed colloids.
      }
  \end{center}
\end{figure}
\begin{equation}
  \label{}
        {\mathcal E}= \sum_{i<j=1}^3{\mathcal E}_{ij}+{\mathcal E}_{123}\,,
\end{equation}
where the first term is sum over the two-body interactions and the second
term ${\mathcal E}_{123}$ is the free energy due to the three body effect. 
For three fixed colloids at the interface, the 
asymptotic three body effect $d_{ij}/R_i \gg 1$ in the limit
$\lambda_{\rm c} \gg R_i, d_{ij}$ reads
\begin{equation}
 \label{eq:three-fixed}
 \frac{{\mathcal E}_{123}}{k_B T} \approx \frac{1}{2} 
 \ln\left[\frac{1-(g_{12}^2+g_{13}^2+g_{23}^2+2g_{12}g_{13}g_{23})}
   {(1-g_{12}^2)(1-g_{13}^2)(1-g_{23}^2)}\right]\,,
\end{equation}
with  $g_{ij} = \frac{\ln(2\lambda_{\rm c}/d_{ij})}
{\sqrt{\ln(2\lambda_{\rm c}/R_{i})\ln(2\lambda_{\rm c}/R_{j})}}$,
$R_i$ the radius of the colloid $i$ and $d_{ij}$ the center-to-center
distance between the colloids $i$ and $j$. {\green Note that the
  energy in Eq.~\eqref{eq:three-fixed} depends only on the distance
  between the colloids.}

For three bobbing colloids at the interface, the 
asymptotic energy in the limit $\lambda_{\rm c} \gg R_i, d_{ij}$ reads
 \begin{equation}
  \label{eq:3-bobing-asym}
  -\frac{\mathcal{E}_{123}}{k_B T} \approx 
  \frac{R_1^4 R_2^2 R_3^2}{d_{12}^4 d_{13}^4}+
  \frac{R_1^2 R_2^4 R_3^2}{d_{12}^4 d_{23}^4}+
  \frac{R_1^2 R_2^2 R_3^4}{d_{13}^4 d_{23}^4}\,.
 \end{equation}
Equation~\eqref{eq:3-bobing-asym} is in agreement with the result in
Refs.~\cite{eft_deserno1,eft_deserno2} and shows that the three body
asymptotic energy is much smaller than the two body interaction $~
d^{-4}$.  Therefore, at large separations, for bobbing colloids the
sum of pairwise energies gives a very good approximation to the 
fluctuation-induced energy.

For three bobbing and tilting colloids, the three body energy in the
limit $\lambda_{\rm c} \gg R_i, d_{ij}$ reads
 \begin{equation}
 \label{eq:3-tilt-asym}
 -\frac{\mathcal{E}_{123}}{k_B T} \approx 81 \left(
 \frac{R_1^8 R_2^4 R_3^4}{d_{12}^8 d_{13}^8}+
 \frac{R_1^4 R_2^8 R_3^4}{d_{12}^8 d_{23}^8}+
 \frac{R_1^4 R_2^4 R_3^8}{d_{13}^8 d_{23}^8}\right)\,.
 \end{equation}
In this case the two body interaction scales with $d^{-8}$, and as
such the pairwise energy summation again provides a good approximation
for the total fluctuation-induced energy.

Figure~\ref{fig:3partfixed} depicts the ratio of the three body
effects to the absolute value of the two body energies vs. the
surface-to-surface separation, $h/R = d_{12}/R-2$, for
two different configurations: colloids sitting on a line
($d_{12}=d_{13}=d_{23}/2$) or at the vertices of an equilateral
triangle ($d_{12}=d_{13}=d_{23}$).  As shown in the figure, there is a
very good agreement between the asymptotic energies (dashed and dotted lines) given by
Eqs.~\eqref{eq:three-fixed}-\eqref{eq:3-tilt-asym} and the numerical results (symbols) calculated by
Eq.~\eqref{eq:energy-final}.

As illustrated in the figure, the three body effect $\mathcal{E}_{123}$ is very small 
at large separations for bobbing only (red circles and triangles) and bobbing and tilting colloids 
(blue squares and diamonds) in both configurations. 
In contrast, the inset in Fig.~\ref{fig:3partfixed} shows that the three body effect energy $\mathcal{E}_{123}$ 
for {\em fixed} colloids is comparable to their two body interaction with the opposite sign,{\it i.e.}, 
there is a strong deviation from the
pairwise summation for the case of fixed colloids in both linear and triangular configurations, see the 
circles and squares in the inset.
Quite interestingly, 
since the three body effect for fixed colloids is positive (repulsive), the total
energy is smaller than the sum of pairwise added interaction energies, {\it i.e.} $\mathcal{E}_\text{tot}<\mathcal{E}_{12}+\mathcal{E}_{13}+\mathcal{E}_{23}$.

In summary, we have presented an effective method to study the
fluctuation-induced forces at fluid interfaces using a modified
scattering method, in which the separation
of the interface and colloid free energies enables us to treat 
various interface and colloid fluctuations. The new approach has
many advantages, such as the treatment of many colloid systems, quick
numerical calculations, and the treatment of different colloid
fluctuations. 

As an example, we studied the three body interaction between frozen
and fluctuating spherical Janus colloids and found a
very interesting difference between them. For both frozen and
fluctuating colloids the two body interactions are attractive. However,
for frozen colloids the three body interactions are repulsive weakening the
interaction compared to the pairwise summation, while for fluctuating colloids the three body
interactions are attractive strengthening the interaction. In addition, 
we found that the three body interaction for the frozen colloids
is non-negligible for the complete range of interactions, while the three
body interactions for the fluctuating colloids are negligible for
larger separations.

Due to the non-trivial behavior of these forces, a
better knowledge of them will shed light on structure formation and
crystallization phenomena at the fluid interfaces.

Authors would like to thank Mehran Kardar for useful discussions.  
This work was supported by the NSF through grants DMR-06-45668 (RZ).

\bibliography{ref.bib}

\end{document}